\documentclass[aps,twocolumn]{aastex631}

\usepackage{graphicx}

\hypersetup{colorlinks = true, allcolors = blue}
\usepackage{amsmath}
\usepackage{longtable}
\usepackage{subfigure}
\usepackage[T1]{fontenc}

\usepackage{parskip}
\shorttitle{A Model-Independent Radio Telescope Dark Matter Search in the L and S Bands}
\shortauthors{Keller et al.}
\graphicspath{{./}{figures/}}

\begin{document}

\vspace{1cm}
\title{A Model-Independent Radio Telescope Dark Matter Search in the L and S Bands}

\author{Aya Keller}
\affiliation{Department of Nuclear Engineering, University of California, Berkeley, CA 94709, USA}

\author{Nicole Wolff}
\affiliation{Department of Physics, Columbia University, NY 10027, USA}


	
\author{Karl van Bibber}
\affiliation{Department of Nuclear Engineering, University of California, Berkeley, CA 94709, USA}

\correspondingauthor{Aya Keller}
\email{ayakeller@berkeley.edu}
\begin{abstract}  
\noindent

Ultralight bosonic dark matter in its most general form can be detected through its decay or annihilation to a quasimonochromatic radio line. Assuming only that this line is consistent with the most general properties of the expected phase space of our Milky Way halo, we have developed and carried out a novel model-independent search for dark matter in the L and S bands. More specifically, the search selects for a line that exhibits a Doppler shift with position according to the solar motion through a static halo and similarly varies in intensity with position with respect to the galactic center. Over the combined L- and S-band range 1020 - 2700 MHz, radiative annihilation of dark matter is excluded above  $\langle\sigma v\rangle \approx 10^{-30} \text{ cm}^3 \text{ s}^{-1}$, and for decay above $\lambda \approx 10^{-32} \text{ s}^{-1}$.

\end{abstract}

\section{Introduction}
While the overwhelming evidence for the existence of dark matter continues to build, progress towards its identification remains minimal. In recent years, advances in detector  technology have enabled a new generation of experiments that are exquisitely sensitive to specific well-motivated dark matter candidates; see \protect\cite{Aprile2019} for WIMP dark matter and \protect\cite{Backes2021} in the realm of ultralight bosonic dark matter. One of these is the axion, an ultralight particle that arises as the solution to the strong-CP problem and whose couplings to Standard Model particles are extraordinarily weak. Axion searches are generally based on their resonant conversion to quasimonochromatic photons in a strong magnetic field, with an expected signal strength on the order of yoctowatts. As the conversion takes place in a high quality-factor microwave cavity, these are narrow-band experiments, which along with the ultraweak signal strength implies years of integration time to cover even an octave of frequency range. While these experiments have been able to successfully exclude regions within a decade of mass, theoretical bounds have recently been relaxed by several orders of magnitude, giving rise to an even more broad range of possible axion masses. Furthermore, attention has recently been shifting to a theoretical framework for dark matter that does not assume specific models. An optimal analysis strategy should thus rely on as few general 
assumptions as possible, while possessing a high degree of selectivity and sensitivity to dark matter.

\section{General Concepts}
This search focuses on the possible radiative decay or annihilation of ultralight dark matter within our Milky Way galactic halo, leading to a quasimonochromatic radio line ($\Delta \nu$/$\nu$ $\approx 10^{-3}$).  This search is further predicated on two generally accepted characteristics of dark matter in our galactic halo. First, we assume that the dark matter constitutes a static halo through which our solar system is moving, with a characteristic velocity $v_S \approx$ 240 km s$^{-1}$ tangential to the disk.  Consequently, such a radio line would be distinguished from any other source, conventional or otherwise, by a systematic Doppler shift with respect to the Sun’s direction of motion: (l, b) = (90$^\circ$,0$^\circ$) in galactic coordinates.  While our galaxy and its halo likely include nonvirialized streams of both baryonic and dark matter due to late-time assimilation of smaller galaxies, overall our halo should be well represented by a static virialized spherical distribution.  Second, the signal should reflect the spatial distribution as represented by a standard halo model; more specifically the signal power should follow the line-integrated density of the halo $\rho$ for dark-matter decay, or $\rho^2$ for annihilation or any other two-body process producing a photon.  This implies that in an all-sky data set, a radio line representing a real dark matter signal should be maximized toward the galactic center, (0$^\circ$, 0$^\circ$), minimized looking outward, and roughly symmetric around that axis.  

\begin{figure}[h]
    \centering
    \includegraphics[width=3cm, angle=90]{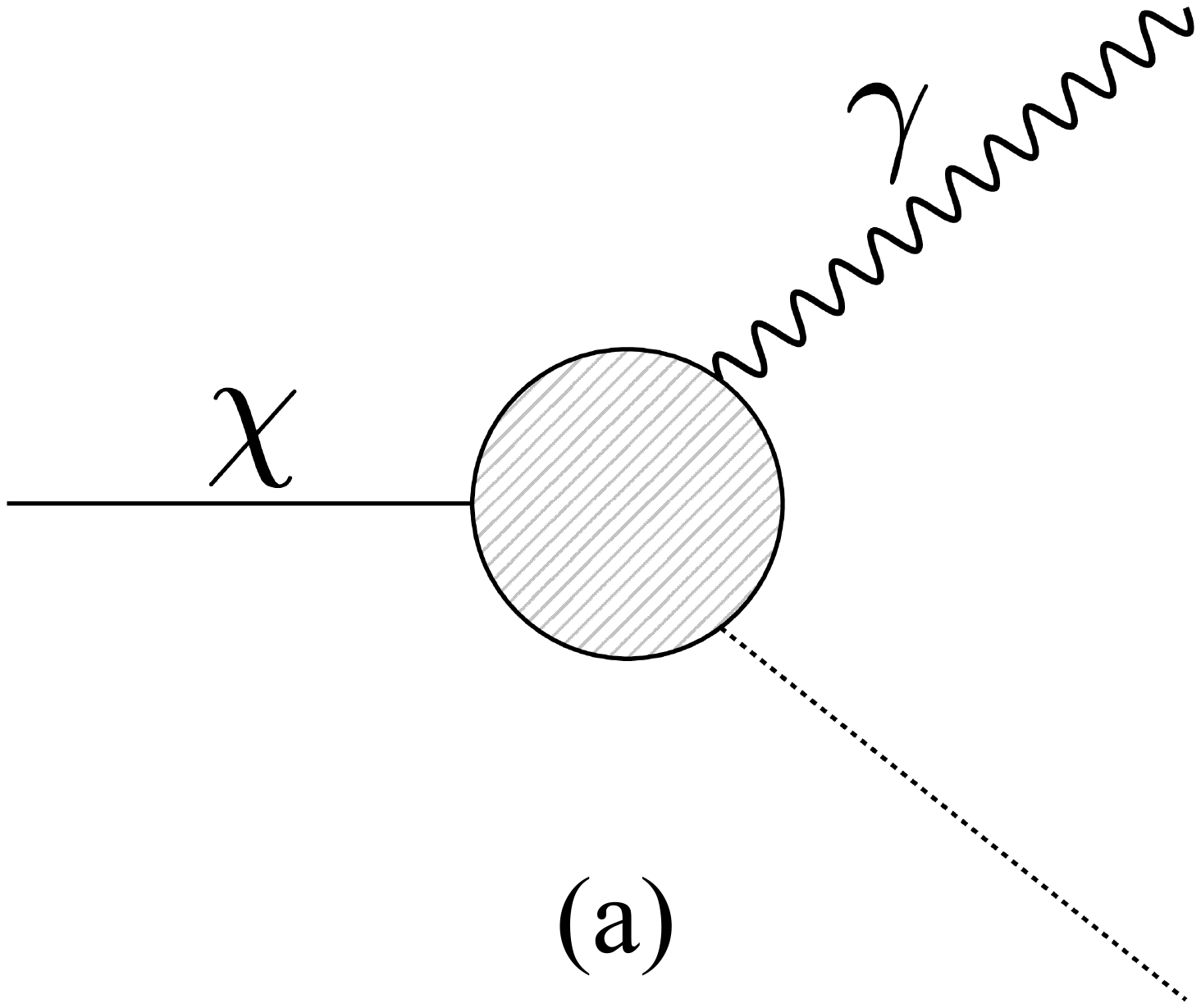} 
    \qquad
    \includegraphics[width=3cm, angle=90]{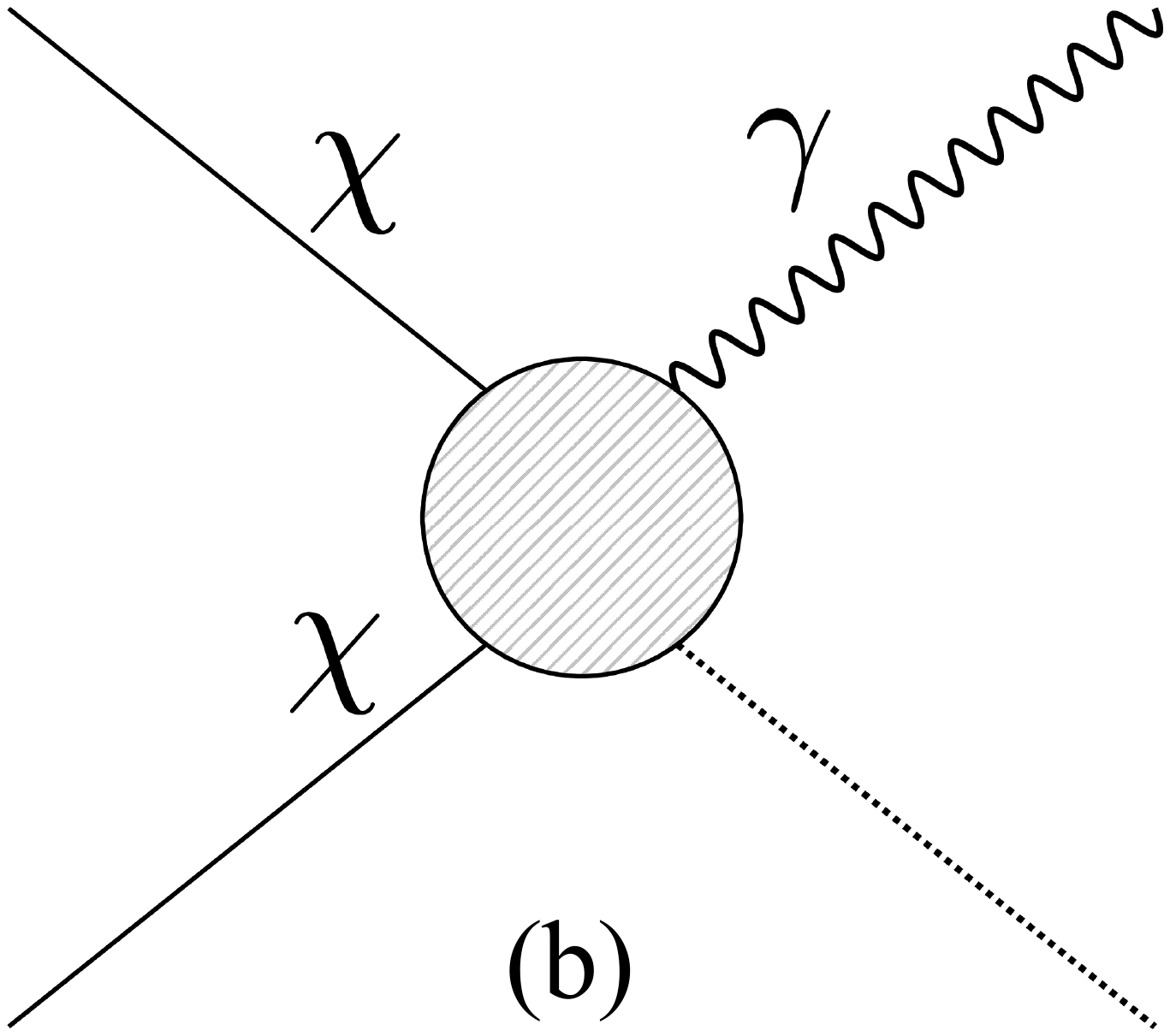} %
    \caption{(a) Decay of a dark matter particle into a photon and another particle or photon.  (b) Annihilation or Compton-like process leading to a two-body final state, one of which is a photon.}%
    \label{fig:example_process}%
\end{figure}

Dark matter decay here includes all processes $\chi \rightarrow \phi + \gamma$, including in principle two-photon decays from  pseudoscalars such as the axion or an axion-like particle $\chi \rightarrow \gamma + \gamma$. For simplicity in calculating and presenting limits on the decay rate, $\lambda$, we will assume the photon energy  $h\nu$ = $\frac{m_\chi c^2}{2}$.  
What we refer to as annihilation includes all two-body initial states, including annihilation proper $\chi + \chi \rightarrow \phi + \gamma$ 
as well as Compton-like processes $\chi + \xi  \rightarrow \phi + \gamma$, where $\xi$ and $\phi$ represent any standard model or beyond-standard model particles; see Figure \ref{fig:example_process}.
\section{The Breakthrough Listen Data Set}
This analysis utilizes the Breakthrough Listen (BL) public data release from the 100 m Robert C. Byrd
Green Bank radio telescope (GBT), which was taken between 2016 January  and 2019 March  (\cite{Lebofsky2019,Price2020}). This data set includes approximately 1700 nearby Hipparcos catalog stars, as well as some targets from other catalogs that were not used for this analysis (\cite{Perryman1997}; \cite{Isaacson_2017}) and 100
nearby galaxies.

The BL observing cadence for each star comprises six 5 minute spectra with the pattern ABACAD, where
A designates an on-target spectrum, with B, C, and D designating different off-target spectra, each a few
degrees away from the target and from one another. All spectra A, B, C, D were included in the analysis, whether catalog stars or dark sky, after carefully examining for any evidence for differences between pairs of spectra, whether on- and off-star, or between pairs of distinct stars.  Such contributions, which might bias the analysis, could be either from conventional astrophysics or extraterrestrial communication, technosignatures etc., as the Breakthrough Listen survey was conducted in support of the SETI program.  We conclude that the catalog stars in this survey were radio-quiet for the purposes of this study, and thus all spectra were included. The BL on-off observing strategy was designed for optimal subtraction of common background. Under the assumption that dark matter is smoothly distributed in the halo and not preferentially localized around any particular star, all
spectra collected are useful to our analysis, independent of specific coordinates.
The spectrograms used in this analysis have an intrinsic channel resolution of approximately 2.8610 kHz and are
imprinted with the polyphase filter-bank structure, a symmetric bandpass function repeated every 1024
channels in the spectrogram data, or approximately 2.93 MHz. The spectra are additionally characterized by a quasi-periodic $\approx$ 15 MHz undulation, of order 10\% in magnitude. As will be described later, to usefully combine and manipulate
thousands of spectra in our analysis requires that both the polyphase filter and undulatory structure be
removed, as well as drifts in both shape and magnitude, evident even between successive 5 minute
observations.
Clearly, the uniqueness of the BL data set for a general dark matter search is the total integration time it
represents: three months of total broadband observation covering the entire sky accessible to the GBT.

\section{Analysis}
This is not the first search for dark matter based on its decay to a quasi-monochromatic radio line (\cite{Blout2001}, \cite{Foster2020}), and future radio searches are under study.  
Our approach differs insofar as it is not restricted to specific particle candidates, specific mechanisms, or to specific astrophysical sites, but is model-independent and utilizes the entire BL data set comprehensively covering the Milky Way. The analysis concept and proof-of-principle appeared in an earlier publication over a very narrow frequency range in the L-band  (\cite{keller}), but as several improvements and optimizations have been made subsequently, the current analysis will be described below.

A high level step-by-step overview may be helpful to understand how limits on dark matter lifetimes and annihilation cross sections are derived.  (i) First, the data is downsampled by a factor of 64. (ii) The individual spectrograms are then unit-normalized by a procedure that preserves the structure of the expected signal width.  (iii) The Doppler asymmetry spectrum is formed based on sorting the overall sample of spectra into Forward and Backward populations, relative to the Sun’s vector velocity in the MW.  The Intensity asymmetry spectrum is similarly formed, relative to the vector to the galactic center.  (iv) For the halo model assumed with its specific parameters, the line integrals and Doppler shifts associated with the galactic coordinates of each spectrum included in the samples above are calculated.  Templates for the Doppler and Intensity asymmetries are calculated specific to the spectra included in the analysis.  (v) The Doppler and Intensity correlation spectra are formed by taking the dot product between their asymmetry spectra and templates at each frequency.  (vi) Finally the Doppler and Intensity correlation spectra themselves are cross correlated, and physics limits are established based on a p-value analysis.

The cases of dark matter decay and annihilation will be treated separately, as their observed line widths will differ.  Assuming a static halo of uniform virial velocity, the Doppler broadening of the source along the line of sight for the two cases is given by 

\begin{equation}
\frac{\sigma_{D}}{\nu} = \frac{\sigma_{\text{vir}}}{\sqrt{3}c} \quad  \text{and} \quad \frac{\sigma_{A}}{\nu} = \frac{\sigma_{\text{vir}}}{\sqrt{6}c} \\
\label{eqn:DopplerBroadening}
\end{equation}
where $\sigma_{D}$ and $\sigma_{A}$ denote the standard deviation in frequency associated with decay and annihilation, respectively, and $\sigma_{\text{vir}}$ the virial velocity of the halo.  Being the sum of Gaussian variates, the distribution of the two-body center of mass motion is narrowed by $\sqrt2$ relative to that of a single particle, and thus the line width for dark matter annihilation is correspondingly narrowed relative to decay.

For each frequency band, this analysis requires the combining of thousands of spectra taken over a year or more, at differing times of day, sky position, atmospheric conditions, hardware and software configurations, etc.  Both the shape and magnitude of the spectra varied by several percent and often much more, even between sequential 5 minute observations taken only a few degrees apart.  To search for a very small signal aggregated over such a large data set first required unit-normalization of the data.  First, the spectra are co-averaged by 64 channels; this results in both improved sensitivity in the final analysis but more importantly a major reduction in computational time in the fitting procedure. The second step in the analysis consists of dividing the spectrograms by the polyphase filter function (PPF).  The deep periodic minima observed in the spectrograms reflect where the neighboring bandpass functions roll off; data points around these minima are characteristically unstable and need to be removed before fitting. 


 At this point, the 2 points at the beginning and end of each of the repeated bandpass responses are interpolated to eliminate residual structure left over from dividing out the PPF. Next, the spectra are unit normalized by forming a ratio of functions fit to the PPF-corrected spectrum  within a window around each data point in the spectrum:

\begin{equation}
R(\nu_i) = \frac{G(\nu_i; \sigma) + P(\nu_i;m, n)}{P_D(\nu_i;m,n)}.
\label{eqn:CorrectionPoly}
\end{equation}

For the polynomial  $P(\nu_i;m,n)$ fit to the data within a window around $\nu_i$, m is the number of data points within the window, and n is the order of the polynomial.  The values $(m,n) = (55, 5)$ were selected on the basis of maximizing the signal to noise of the spectral asymmetry for synthetic signals injected into the spectrum, as described below. For the Gaussian function $G(\nu_i; \sigma)$, $\sigma$ is specified based on the expected width of the signal at  $\nu_i$, and the coefficient and center are left as free parameters for the fit (with the condition that the coefficient is positive). The polynomial in the denominator, $P_D(\nu_i;m,n)$, is the same functional form as that in the numerator, but where a window representing the expected signal width around the center of the Gaussian is blanked out for the fit. 

\begin{equation}
G(\nu_i;\sigma) = a^2 \cdot \exp\left[\frac{-(\nu-\nu_{i})^2}{2\sigma^{2}}\right]
\label{eqn:Windowfunc}
\end{equation}
\begin{equation}
P(\nu_i;m,5) = b + c\cdot \nu_i + d\cdot\nu_i^2 + f\cdot\nu_i^3 + g\cdot\nu_i^4 + h\cdot\nu_i^5
\end{equation}

\begin{figure}[h]

    \centering
    \hspace{.1cm} \includegraphics[width=1\linewidth]{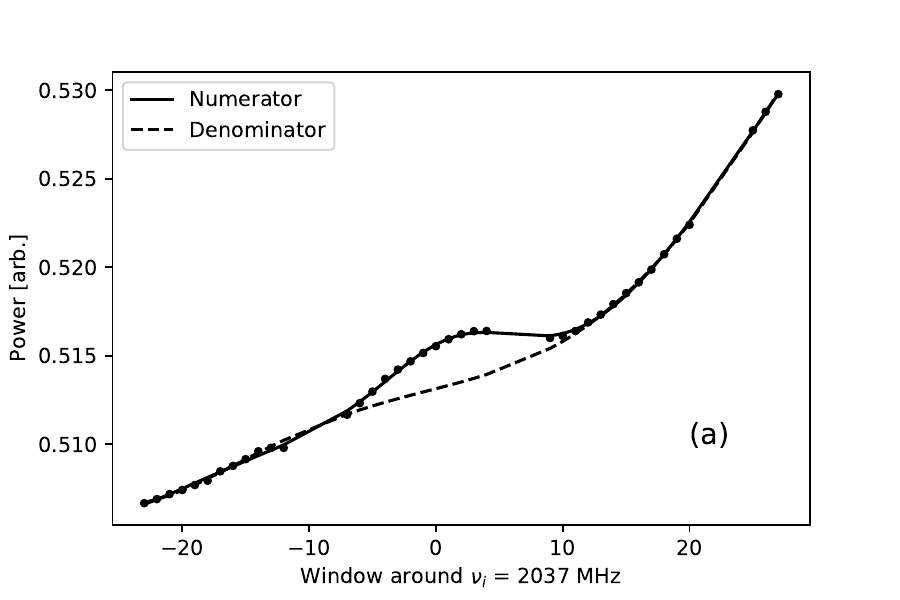}
    \qquad
    \includegraphics[width=1\linewidth]{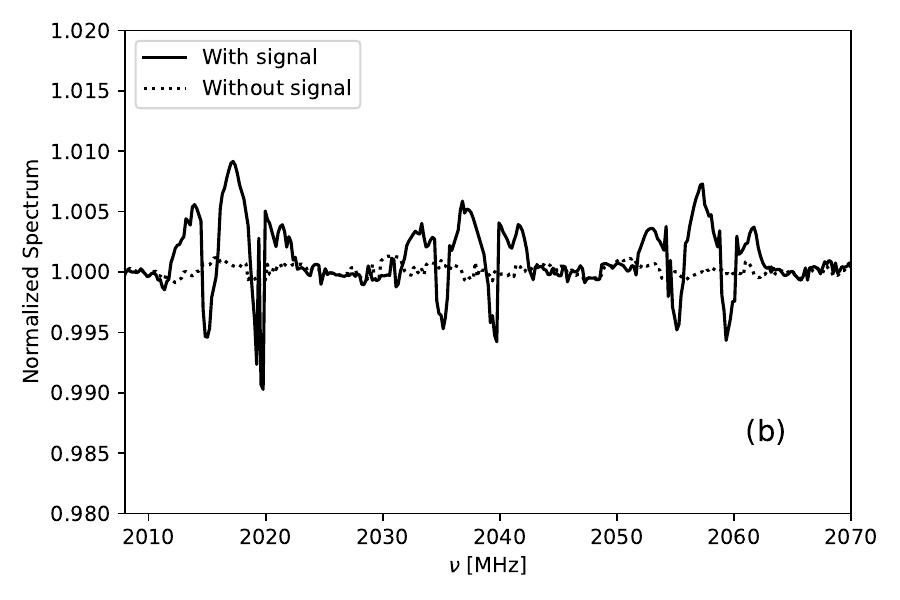} %
    \caption{(a)  A synthetic signal from dark matter decay with  decay constant $\lambda$ = $1 \times 10^{-30}\text{ s}^{-1}$ injected into one raw spectrogram, with the corresponding weighted and unweighted polynomial fits.  (b)   A normalized spectrum with 3 injected signals spaced 20 MHz apart.}%
    \label{fig:example_norm}%
\end{figure}

This procedure ensures that the sum of the polynomial and Gaussian in the numerator more closely tracks any signal atop the local background, whereas the de-weighted polynomial in the denominator is desensitized to any putative signal around the center of the signal found by the fit and thus more closely interpolates the real background in the absence of such a signal. Figure \ref{fig:example_norm} shows both the unweighted and weighted polynomials in the vicinity of an injected decay signal for a decay constant $\lambda$ = $1 \times 10^{-30}\text{ s}^{-1}$ and the resulting normalized spectrum both with and without the injected signal.  

\begin{figure*}[hbt!]
	\centering
	\includegraphics[width=.9\textwidth]{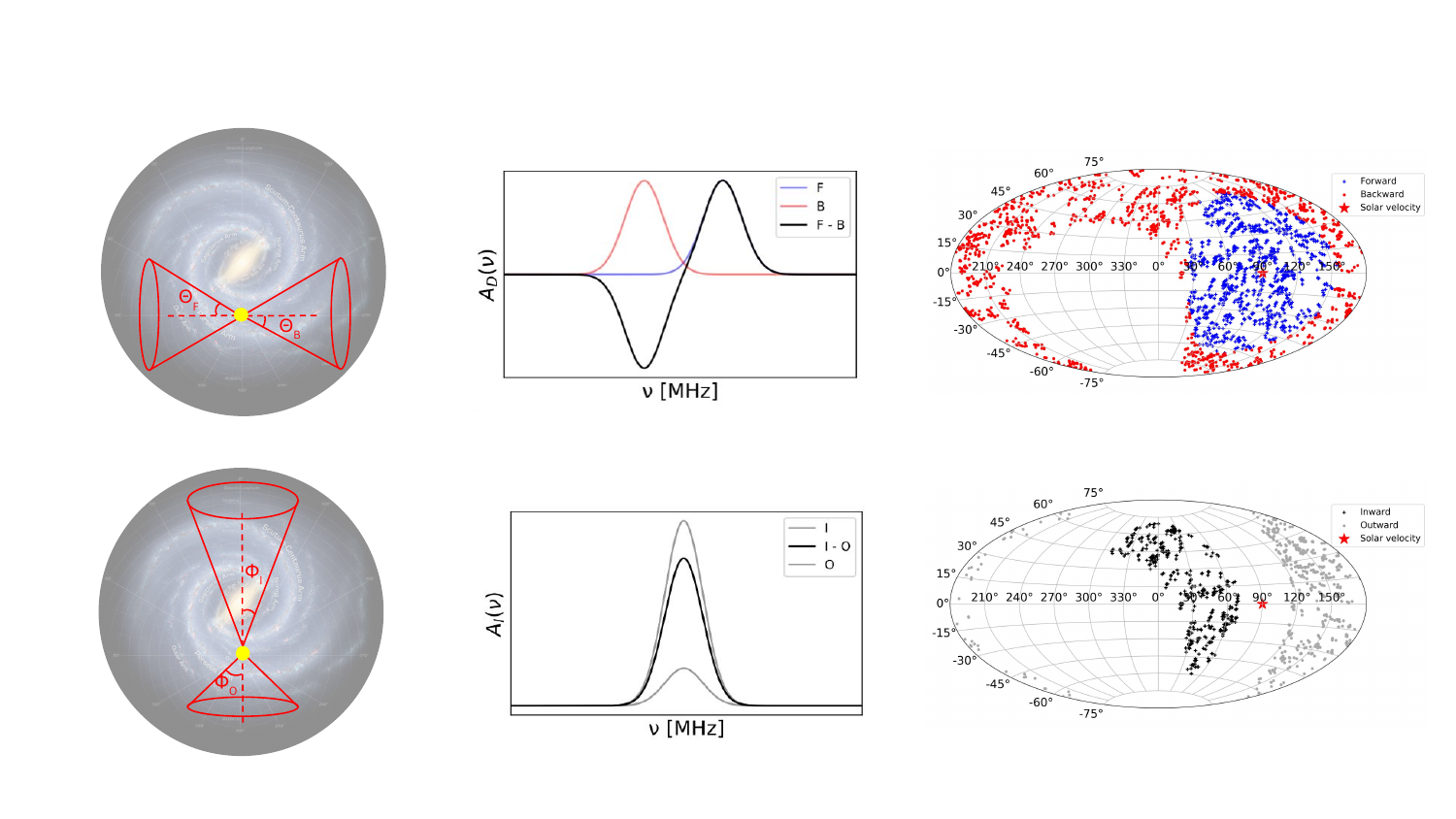}
	\caption{(a) Concept of asymmetry-based searches for dark matter within a large data set broadly sampling the observable sky. The angles defining the Forward and Backward samples, $\theta_{F,B}$ are chosen to maximize the signal-to-noise ratio. (b)  The idealized dark matter signature in the Doppler asymmetry spectrum $A_D(\nu)$ would be a bipolar signal in frequency, prescribed by the $(V_{S},\sigma_\text{vir})$ and the ensemble of specific targets selected.  The idealized dark matter signature in the Intensity asymmetry spectrum  $A_I(\nu)$ would be a unipolar signal centered at the frequency of the decay or annihilation photon in the dark matter’s rest frame. In practice, due to the normalization scheme, the actual asymmetries are more complex in shape, but remain antisymmetric and symmetric respectively. (c) Mercator plots exhibiting the actual targets included in the analysis, the regions being demarked by $\theta_{F}$ = $\theta_{B}$ = $65^{\circ}$ \text{and} $\theta_{I}$ = $70^{\circ}$, $\theta_{O}$ = $115^{\circ}$.
	\label{fig:galacticmap}}
\end{figure*}

Two points should be noted about the normalization method adopted for this analysis.  First, as with many such schemes, the fitting procedure transforms a simple signal shape (here a Gaussian) superposed on a 
smooth background into a more complex form.  This does not represent a fundamental problem, as the overall analysis can be adequately modeled.   Second, as the normalization scheme is optimized to accentuate the signal, it also accentuates the residual structure associated with imperfect removal of the polyphase filterbank bandpass response, as in this frequency range, they are of the same order in width, i.e. a few MHz.  This is a more significant problem, but which the asymmetry analysis significantly cancels out.

The concept of the Doppler and Intensity asymmetries is schematically represented in Figure \ref{fig:galacticmap}.  In the case of the Doppler asymmetry, under the assumption of a static halo, a dark matter signal is Doppler shifted according to its polar angle $\theta$ from the direction of the Sun’s motion through the galaxy, 
\begin{equation}
\nu' = \nu \left(1+\frac{V_S}{c} \cdot \cos \theta \right)
\label{eqn:dopplershift}
\end{equation}

where estimates for the solar velocity $V_S \approx$  240 km $\text{s}^{-1}$ (\cite{Monari2018}) are comparable to the local virial velocity of the galaxy, $\sigma_\text{vir} \approx$  270 $\text{km } \text{s}^{-1}$ (\cite{Pillepich2014}).  Owing to the observational and modeling uncertainties intrinsic to these numbers, we explored the variation of our analysis in excursions of $V_S$ and $\sigma_\text{vir}$ within the range 225 – 275 $\text{km } \text{s}^{-1}$ and find it to be not highly sensitive.

For each case, the asymmetry spectrum is formed:

\begin{equation}
A_{D}(\nu) = \frac{F-B}{F+B} \quad A_{I}(\nu) = \frac{I-O}{I+O} 
\label{eqn:aysequations1}
\end{equation}
with F (B) designating the average of all spectra within the Forward (Backward) acceptances defined by their polar angles $\theta_{F/B}$.

\begin{equation}
F(\nu, \theta_{F}) = \frac{1}{n_f} \sum_{i} f_i(\nu) \quad B(\nu, \theta_{B}) = \frac{1}{n_b} \sum_{i} b_i(\nu)
\label{eqn:aysequations2}
\end{equation}
for some fixed $\theta_{F}$ and $\theta_{B}$ and similarly for I (O), the Inward (Outward) populations within their respective polar angle cuts $\Phi_{I/O}$.  Forming asymmetry spectra has the virtue of canceling out to a high degree common-mode residual structure apparent in all normalized spectra at the $\approx 10^{-4}$ level, thus enabling a more sensitive search. 

We wish to note that this is not the first proposed use of a Doppler shift as a discriminant of dark matter; see \cite{Speckhard} for its suggested application in regard to the 3.5 keV line reported by several X-ray observatories.

\subsection{Doppler asymmetry analysis}
The flux density in general for the two cases of annihilation and decay processes is given by:
\vspace{.5cm}
\begin{equation}
\frac{P_{A}}{\Delta A \Delta \nu} =\frac{1}{8\sqrt{2\pi}}\frac{\langle\sigma v\rangle (\Delta \theta)^2 c^2}{M_{\chi}\eta^{A} \nu_0}  e^{\left[-(\frac{\nu-\nu_0}{\sqrt2 \eta^{A} \nu_0})^2\right]} \int_{0}^{\infty}\rho(\mathbf{r})^{2} \,d\mathbf{r} \\
\label{eqn:Annilhationpower}
\end{equation}

\begin{equation}
\frac{P_{D}}{\Delta A \Delta \nu} =\frac{1}{16\sqrt{2\pi}}\frac{\lambda (\Delta \theta)^2 c^2}{\eta^{D} \nu_0}  e^{\left[-(\frac{\nu-\nu_0}{\sqrt2 \eta^{D} \nu_0})^2\right]} \int_{0}^{\infty}\rho(\mathbf{r}) \,d\mathbf{r} \
\vspace{.4cm}
\label{eqn:Decaypower}
\end{equation}
which depends on the particle physics through the velocity-weighted cross section $\langle\sigma v\rangle$  for annihilation, or the decay constant $\lambda$  in the case of decay. Here  $\eta$  is the line width with $\sigma_\text{vir}$ the halo viral velocity, $\rho(\mathbf{r})$  the halo density along the line of sight, and $\Delta \theta$ the frequency-dependent FWHM beamwidth of the telescope.  

\begin{equation}
\eta^A = \frac{\sigma_\text{vir}}{\sqrt{6}c}, \ \ \  \eta^D =  \frac{\sigma_\text{vir}}{\sqrt{3}c}  \\.
\label{eqn:linewidthequ}
\end{equation}





For simplicity, these formulae are written for the special case of a two-photon final state, i.e. $\chi \chi \rightarrow \gamma\gamma$   and $\chi \rightarrow \gamma\gamma$, for which the resulting limits will be seen to be much weaker than allowed by stellar evolution but are readily convertible to limits for final states with a massive particle $\phi\gamma$.

The resulting limits are derived by a standard matched-filtering technique.  A template of the asymmetry for a dark matter signal is created at each frequency $\nu$ in the spectrum, $\textbf{T}(\nu)$, specific to the particle physics input $\langle\sigma v\rangle$ or $\lambda$, target samples $\theta_{F/B}$, galactic parameters $(V_{S},\sigma_\text{vir})$, and halo model.  This involves calculating the contribution of the expected signal for each target (Eqs. \ref{eqn:Annilhationpower},\ref{eqn:Decaypower}), transforming the input Gaussian line into the actual shape as it appears in the normalized spectrum, applying the coordinate-dependent Doppler shift (Eq. \ref{eqn:dopplershift}), and finally summing and forming the asymmetry (Eqs. \ref{eqn:aysequations2}, \ref{eqn:aysequations1}).  At each frequency $\nu$ the template is integrated over the asymmetry spectrum:

\begin{equation}
R_D (\nu) =  \boldsymbol{T} \cdot \boldsymbol{A_D(\nu)}  \equiv  \sum_{\nu'}T(\nu - \nu') A_{D} (\nu - \nu')
\label{eqn:dopplerspectrum}
\end{equation}
resulting in the Doppler correlation spectrum $R_D(\nu)$.  

\begin{figure}[hbt!]
    \centering
    \includegraphics[width=.8\linewidth]{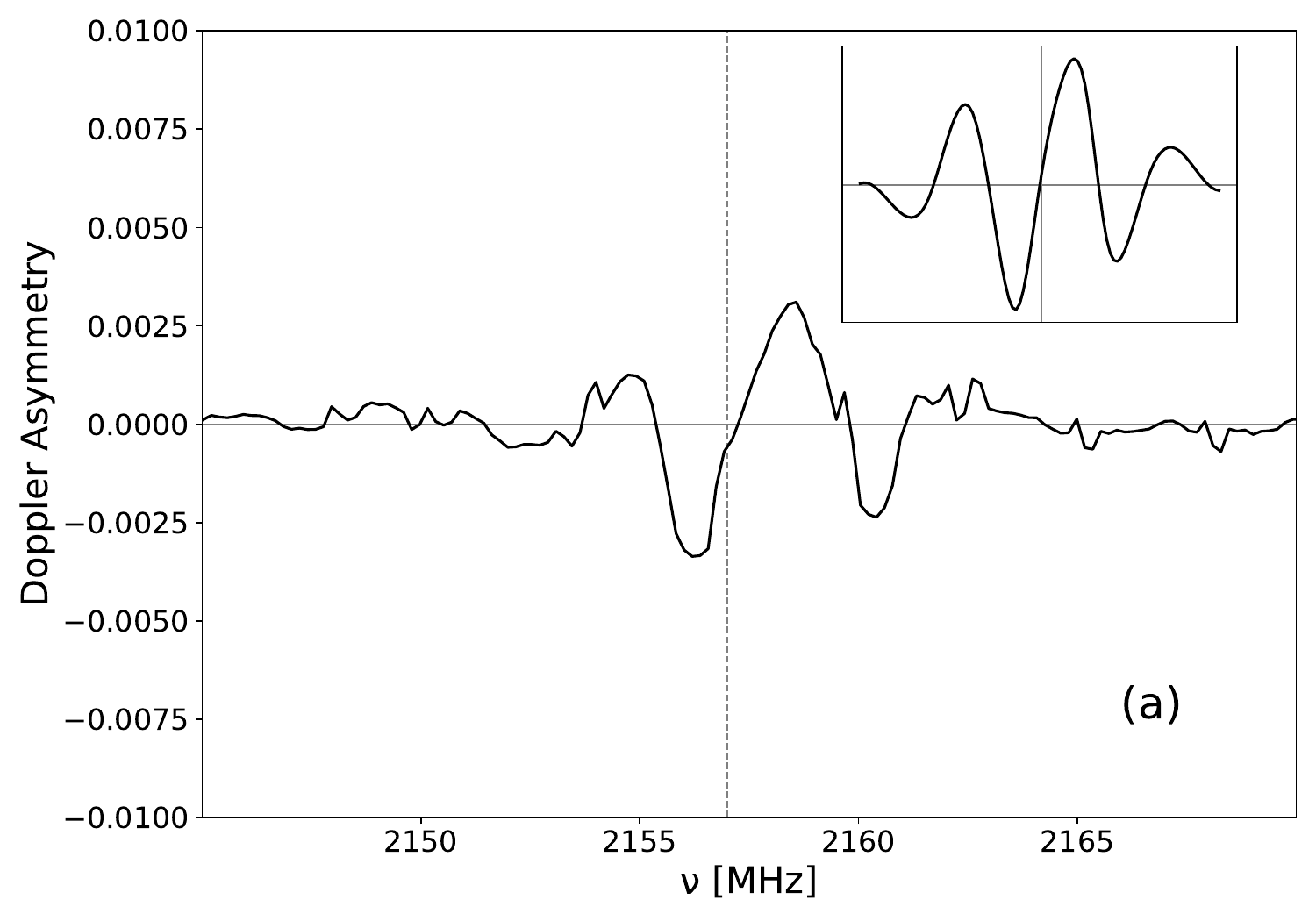} 
    \qquad
    \hspace{.05cm}\includegraphics[width=.8\linewidth]{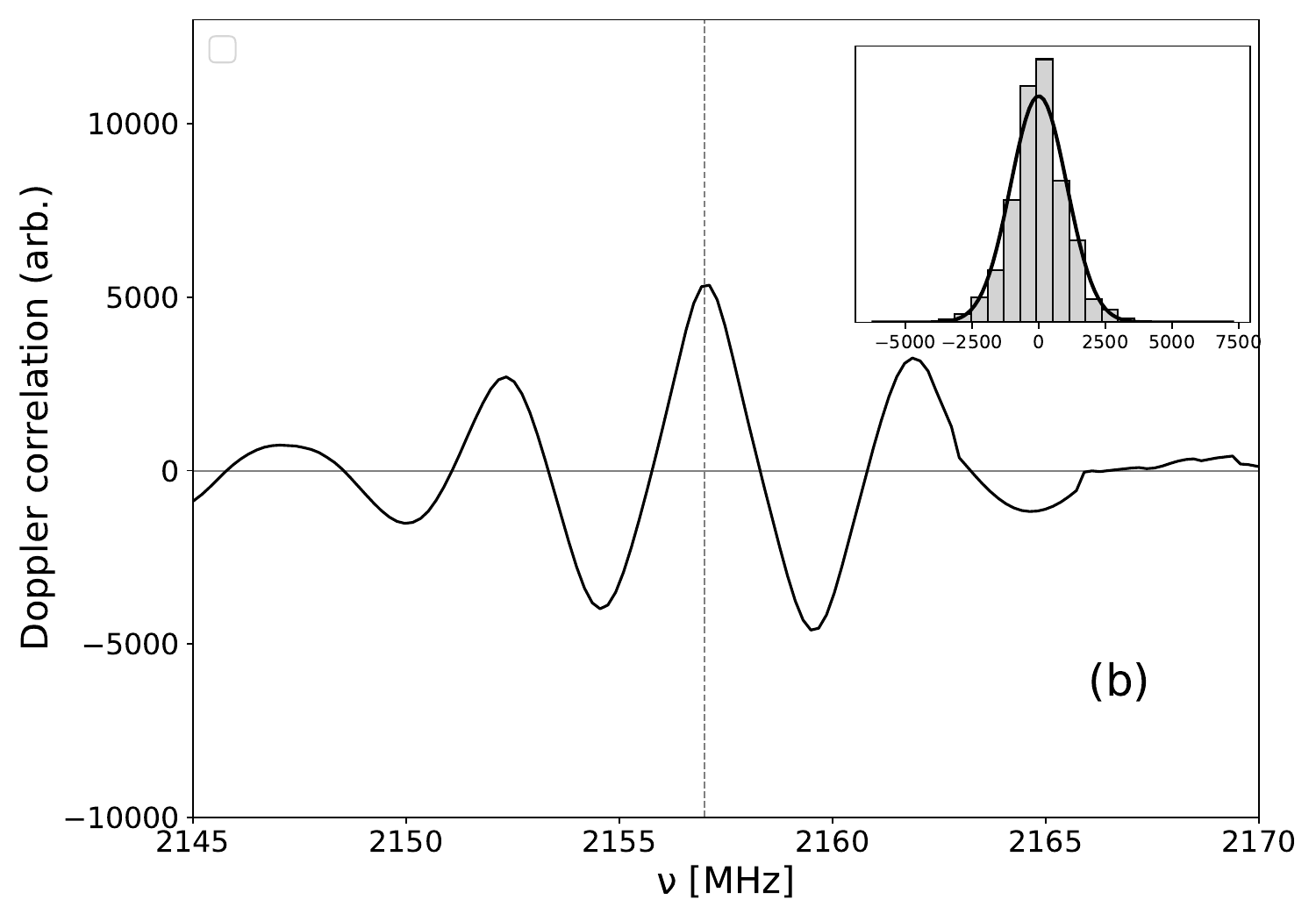} %
    \caption{(a)  Doppler asymmetry spectrum for the case $V_S$ = 225 km $\text{s}^{-1}$, $\sigma_\text{vir}$ = 250 km $\text{s}^{-1}$, for $\nu$ = 2157 MHz, and a decay constant $\lambda$ = $1 \times 10^{-30}\text{ s}^{-1}$ shown for scale. (Inset) Template formed for current parameters. (b) The Doppler correlation spectrum.  (Inset)  Doppler correlation spectral distribution in units of standard deviation $\sigma_\text{corr}$ in the absence of a signal.}%
    \label{fig:example_dopp}%
\end{figure}

Limits are derived by injecting synthetic signals at the raw spectrogram level, carrying through the analysis as described, and establishing the confidence level against the statistical distribution of $R(\nu)$ in the absence of a signal.   The signal-to-noise ratio (S/N) is maximized by utilizing all of the data after the data quality cut, and equalizing the Forward and Backward populations, corresponding to the polar angle $\theta_F$ = $\theta_B$ = 65$^{\circ}$ (Figure \ref{fig:example_dopp}).

\subsection{Intensity Asymmetry Analysis}

The analysis based on the expected signal asymmetry between looking inward (I), toward the galactic center and outward (O), away from the galactic center follows in an essentially identical manner.  As mentioned previously, here the S/N is maximized by more tightly restricting the cone defining the inward population, and leaving a large gap in angle ($\theta_I$ = 70$^{\circ}$  $\theta_O$ = 115$^{\circ}$). For this first analysis, a Navarro-Frenk-White (NFW) halo was used~(\cite{Navarro1997,Nesti2013}):

\begin{equation}
\rho (r)  =  \rho_c \left(\frac{r}{r_c}\right)^{-1}\left(1+\frac{r}{r_c}\right)^{-2}
\label{eqn:NFWprofile}
\end{equation}
with $\rho_c = 1.4 \times 10^{7} \space \text{ M}_{\odot} \text{kpc}^{-3}$ and $r_c$ = 16.1 kpc.

\begin{figure}[h]
    \centering
    \includegraphics[width=.5\textwidth]{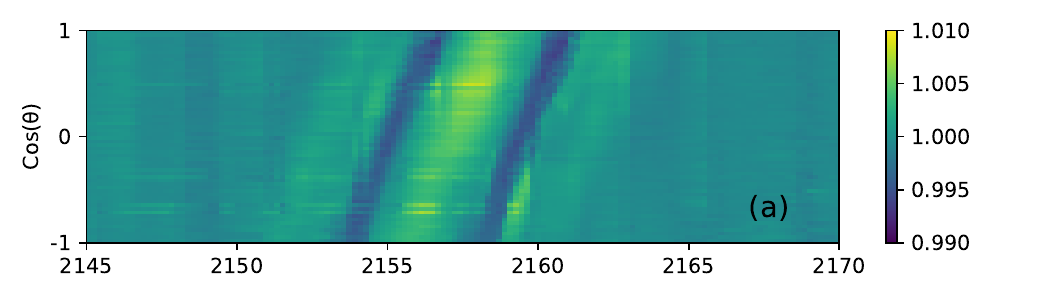}
    \includegraphics[width=.5\textwidth]{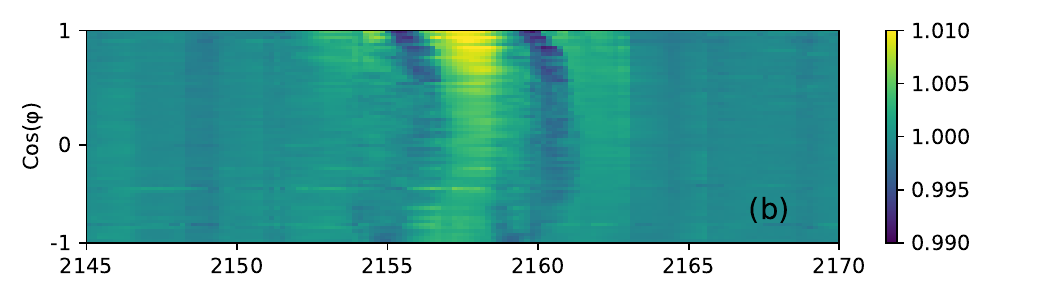}
   
    \caption{False-color maps of normalized spectra sorted by features of the data corresponding to the Doppler and Intensity analyses: (a) $\theta$ (angle from the sun's direction of motion), $\phi$ (angle from the galactic center) for a signal of size $\lambda$ = $1 \times 10^{-30}\text{ s}^{-1}$. \label{fig:heatmaps}}
\end{figure}

\begin{figure}[hbt!]
    \centering
    \includegraphics[width=.8\linewidth]{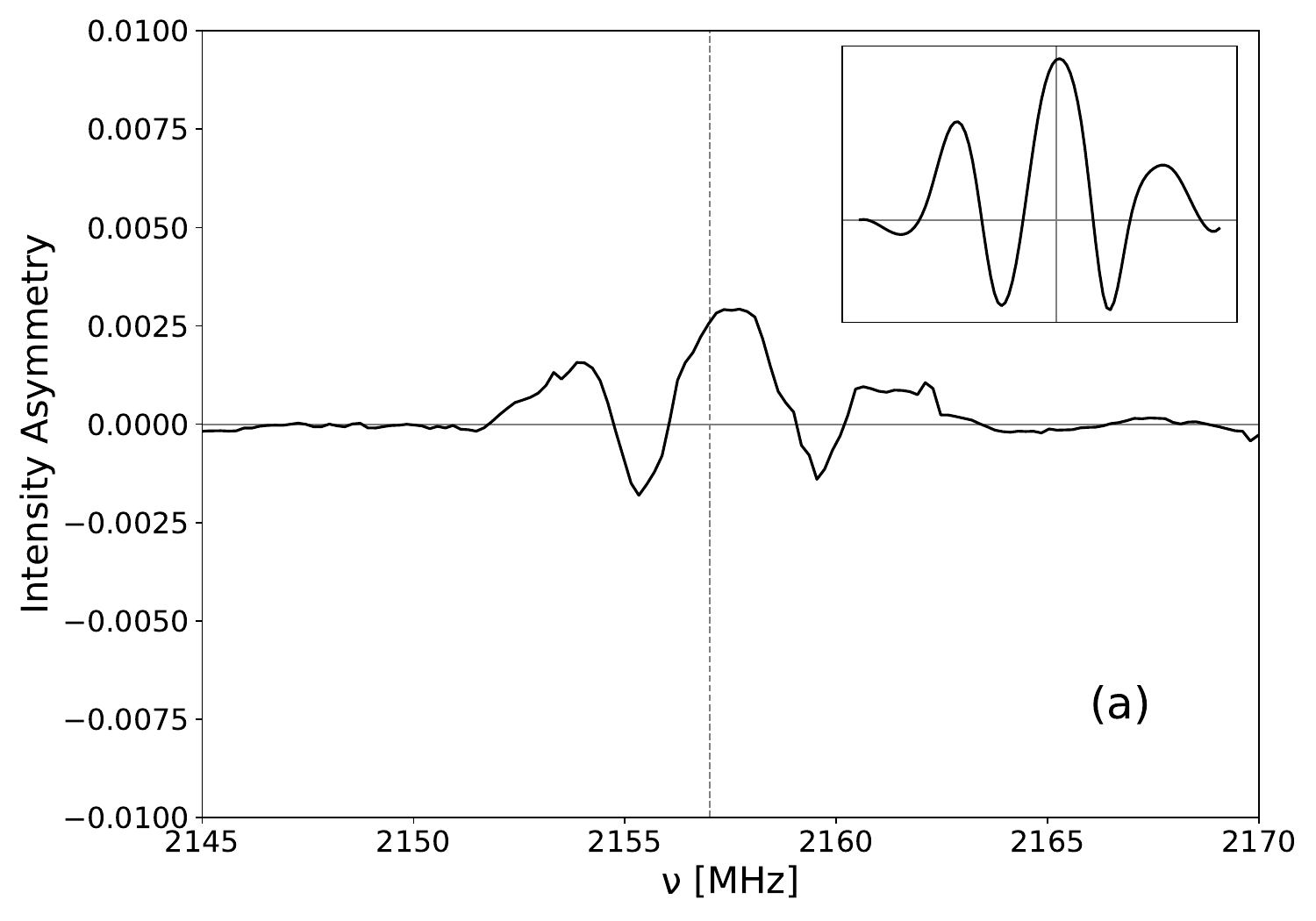} 
    \qquad
    \hspace{.05cm}\includegraphics[width=.8\linewidth]{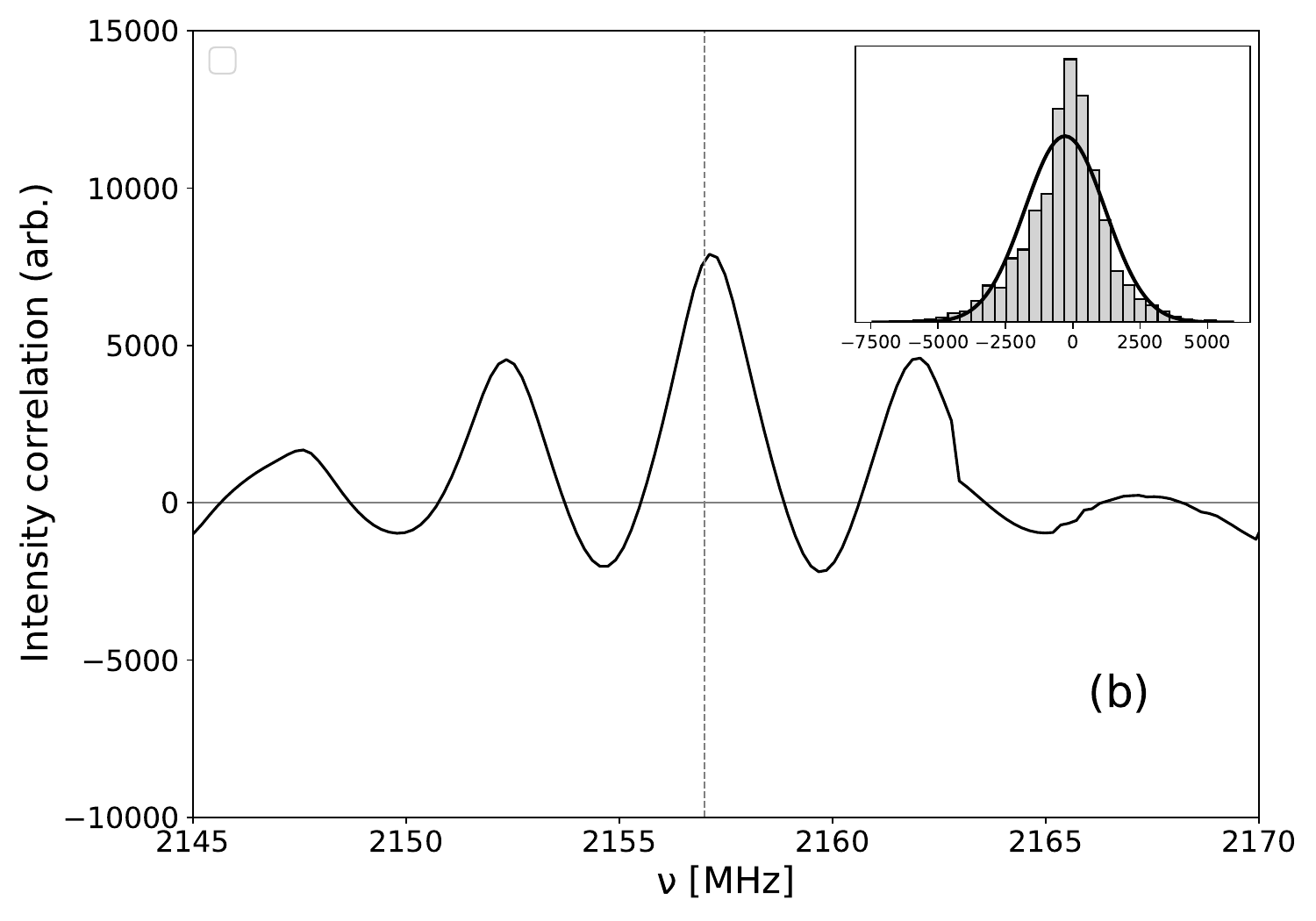} %
    \caption{(a)  Intensity asymmetry spectrum.  (b) Intensity correlation spectrum and its spectral distribution, for the same case as in Figure 5.}%
    \label{fig:example_Intensity}%
\end{figure}

Stronger limits can be achieved by cross-correlating the Doppler and Intensity analyses, ensuring that the two analyses are absolutely independent.  In particular, it is most important to ensure that there are no remnant velocity correlations inadvertently built in to the I,O populations of the Intensity analysis, owing to the incomplete GBT galactic coverage (Figure \ref{fig:galacticmap}).  To ensure that the total I,O contributions (Eq. \ref{eqn:aysequations2}) were coincident in frequency, that is, free of residual Doppler shift, while maximizing the number of spectra selected, the Hungarian matching algorithm is used for annihilation and a manual shift of the signal is used for decay as the decay signal is too wide to necessitate the matching (\cite{Kuhn1955}). Figure \ref{fig:example_Intensity} shows the corresponding Intensity asymmetry $A_I(\nu)$ and Intensity correlation $R_I(\nu)$.
\subsection{Combined Analysis}
The results of the separate Doppler and Intensity analyses can be combined to yield stronger limits on halo dark matter.  As the analyses are independent and their individual template correlation spectra are approximately Gaussian distributed, application of the p-test on the cross correlation of the two analyses is straightforward. 
\begin{figure}[h]
    \centering
    \includegraphics[width=.4\textwidth]{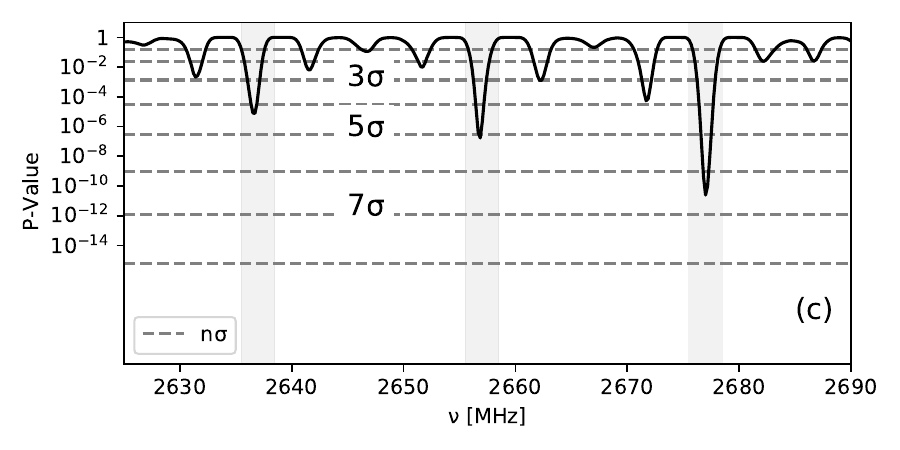}
    \hspace{.05cm}
    \includegraphics[width=.4\textwidth]{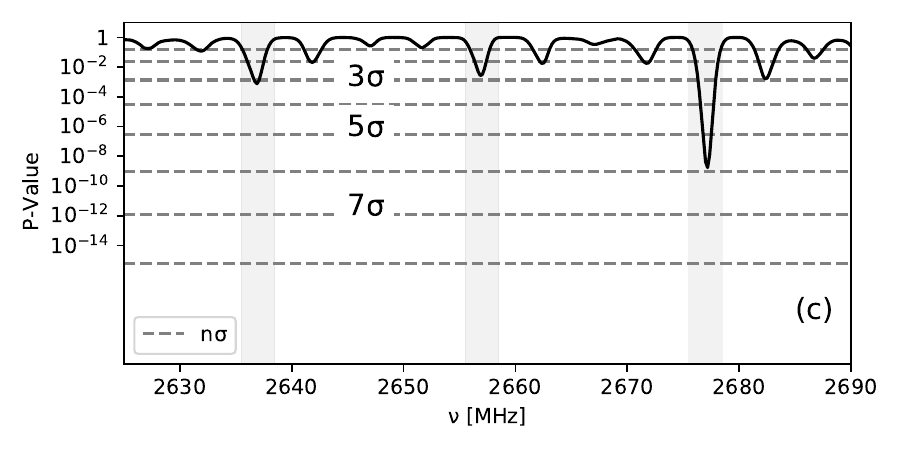}
     \hspace{.05cm}
    \includegraphics[width=.4\textwidth]{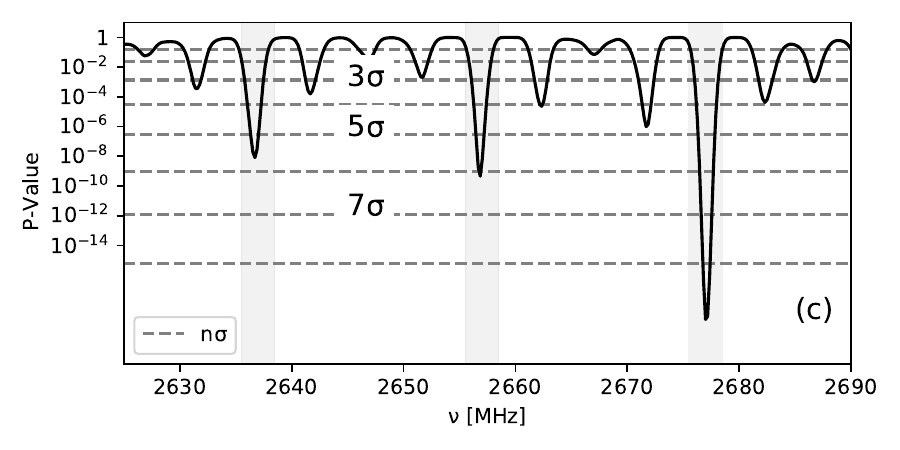}
    \caption{The p-statistic for 3 injected signals over a clean region in the S Band, for the decay case; $\lambda$ = $1 \times 10^{-30}\text{ s}^{-1}$. (a) Doppler p-value spectrum, (b) Intensity p-value spectrum, (c) combined p-value spectrum.\label{fig:pvalue}}
\end{figure}

Three comments are in order concerning determination of the limits. 
First, as the BL data set was lacking in any useful calibration targets, the system equivalent flux density (SEFD) was calculated from the published GBT L- and S-band system noise temperature of T$_{\text{SYS}}$ = 20 K and aperture efficiency 70\%. To check our absolute calibration, a comparison was made of the inferred HI column density measured in our data with data from the Bonn Library (\cite{Bekhti2016}).   The inferred value is within 1.5\% of the older Leiden/Argentine/Bonn (LAB) survey and 15\% higher than the newer Effelsburg/Bonn HI survey (EBHIS), thus giving us confidence in our procedure.
 Second, the limit above was derived with a search template of the same galactic halo parameters as the injected signal, ($V_S$, $\sigma_\text{vir}$) = (225, 250) [km s$^{-1}$].  As we have no precise à priori knowledge of the halo parameters, we tested 2 additional cases in the L Band within current bounds of solar and virial velocities,  ($V_S$, $\sigma_\text{vir}$)= (225, 250), (200, 250), (225, 275) [km s$^{-1}$]. By mismatching the injection and search parameters, we measured the sensitivity of our search to our ignorance of these parameters. 
 \begin{figure}[h]
    \centering
    \includegraphics[width=1\linewidth]{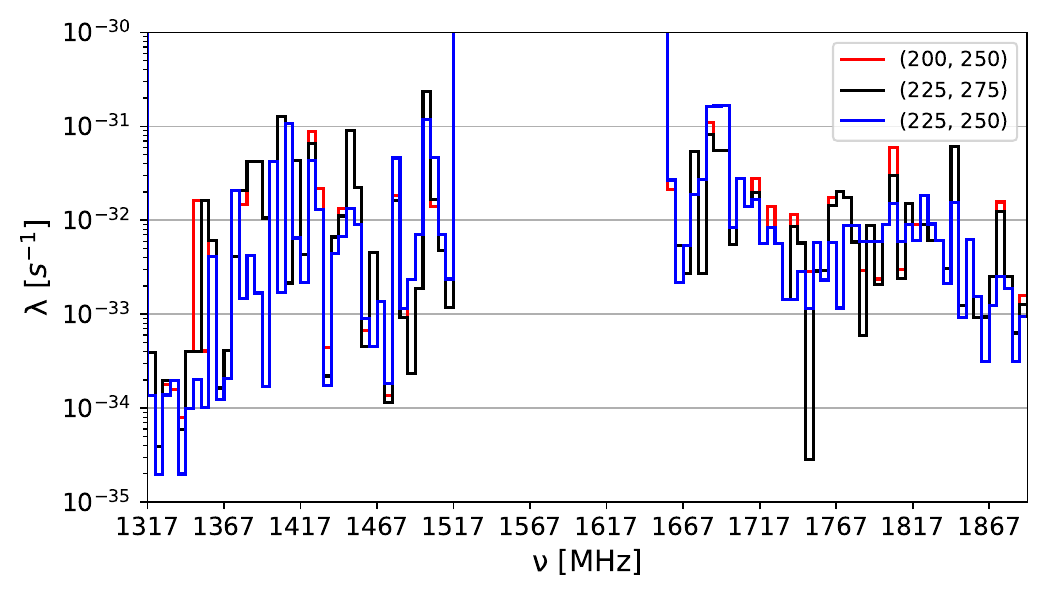} 
   
    \caption{Decay limits across the L band for 3 cases of the injection parameters (V$_{solar}$, V$_{virial}$). For all cases, the search parameters were (225, 250).}%
    \label{fig:3lims}%
\end{figure} 
As shown in Figure \ref{fig:3lims}, the resulting limits differ by about 10\% on average. Third, it may be inquired to what degree the physics exclusion limits depend on the exact choice of halo model.  We have performed such a comparison between those derived from the NFW halo described above, and those from a Burkert halo, parameters of which were determined by fitting to the same Milky Way galactic observables by the same authors (\cite{Nesti2013}).  Whereas the NFW represents a cuspy halo, divergent at the galactic center, the Burkert model represents a cored halo with a central density of $4.13 \times 10^{7} \text{  M}_{s} \text{ kpc}^{-3}$, where $\text{M}_{s}$ is the solar mass.  While both yield essentially the same local density, the quantity of interest to direct detection experiments, $\rho^\text{NFW} (r_s) = 0.471 \text{ GeV} \text{ cm}^{-3}$ and $\rho^{Bur} (\text{r}_s) = 0.487 \text{ GeV} \text{ cm}^{-3}$, their virial masses $\text{M}_\text{vir}$ differ considerably,  $1.53 \times 10^{12} \text{  M}_{s}$  and  $1.11 \times 10^{12} \text{  M}_{s}$ respectively.  As our radio survey method is sensitive to both the halo distribution and its total mass, it is not surprising then that the physics limits will vary with the choice of halo model and in fact the limit on the annihilation cross section for the Burkert model at 1775 MHz being weaker by 60\%. Analogous uncertainties in direct detection limits arise from differing assumptions of local halo density, which varies among reports from $0.3 \text{ to } 0.6 \text{ GeV} \text{ cm}^{-3}$. 
\section{Results}
\begin{figure*}[hbt!]
	\centering
	\includegraphics[width=1\textwidth]{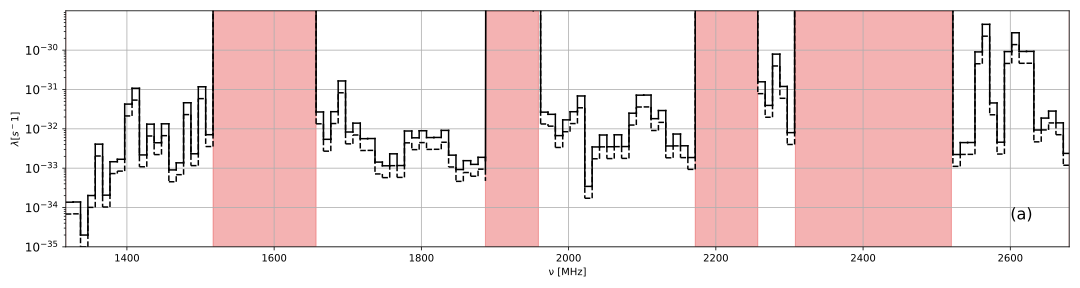}
    \includegraphics[width=1\textwidth]{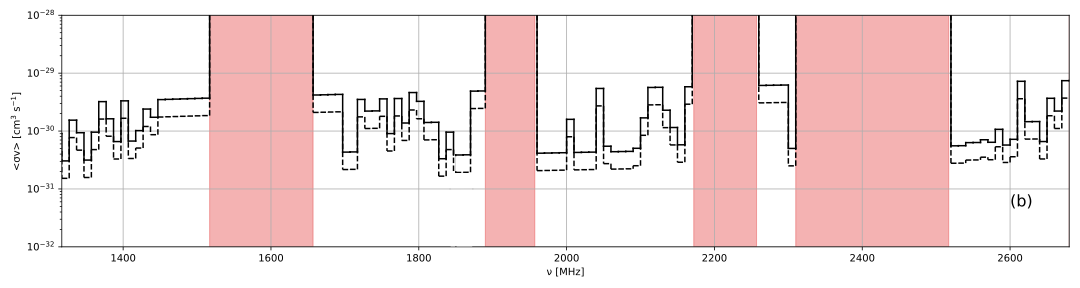}
	\caption{(a) Decay limits in the L and S Bands, with excluded regions highlighted in red (for reasons of high noise or damping).(b) Annihilation limits in the L and S Bands. This figure corrects the absolute value of the velocity-weighted cross section appearing in \protect\cite{keller} for the frequency range 1710-1840 MHz. In both cases, the solid line represents 1 photon final states and the dashed line represents 2 photon final states. \label{fig:an_limits}}
\end{figure*}

Resulting limits of annihilation and decay are shown in Figure \ref{fig:an_limits}. These were calculated by reducing the injected signal until the resulting combined p-value reached 3$\sigma$ for each frequency window of 5 MHz. Any other candidates above 3$\sigma$ were excluded by investigation of RFI, temporal correlations, and lack of a systematic Doppler shift.

These limits also include the effect of stimulated emission. Caputo et al. have calculated the stimulated emission of axions within the radiation background of our galactic halo~(\cite{Caputo2019}), but the results are readily generalized to any process with a photon in the final state.  Specifically, the enhancement for the rate or cross section is a simple multiplicative factor of $2f_{\gamma}$, or $f_{\gamma}$ if there is only a single photon in the final state, where  $f_{\gamma}$ is the photon occupation number.
For our halo, 
\begin{equation}
f_{\gamma}  = f_{\gamma,CMB} (\nu) + f_{\gamma,ext-bkg} (\nu) + f_{\gamma,gal} (\nu;l,b) 
\label{eqn:galatichalocontributions}
\end{equation}
where the three components represent the cosmic microwave background ($T_\text{CMB} = 2.725 K$), the extragalactic radio background ($T_\text{ext-bkg} \approx 0.78 K$  in the middle of our frequency range), and the galactic diffuse emission.  The CMB and the extragalactic radio background are isotropic, whereas the galactic diffuse emission is sharply peaked around the galactic center, which being largely below the GBT horizon, can be neglected for this analysis.

\section{Discussion}
The analysis above has been carried out assuming two-photon final states, and only for annihilation in the case of two-body initial states, which are of equal mass by construction.  The results are readily generalized when deviating from these assumptions.  In regard to the frequency dispersion of the search template, whereas for annihilation one expects a narrowing relative to the decay width of $\sigma_{A} = \sigma_{D} / \sqrt{2}$, for the asymmetric case $m_{\xi} \neq m_{\chi}$,

\begin{equation}
\frac{\sigma_{D}}{\nu} = \frac{\sigma_{V}}{\sqrt{3}c} [(\frac{m_{\chi}}{M})^2+(\frac{m_{\xi}}{M})^2]^{\frac{1}{2}}
\label{eqn:decaywidthcontributions}
\end{equation}
with $M = m_{\xi} + m_{\chi}$; note the bracketed factor is $\frac{1}{\sqrt{2}}$ for $m_{\chi} = m_{\xi}$, but approaches 1 for $m_{\xi} << m_{\chi}$.  Thus anticipating all possibilities, including e.g. a Compton-like process involving an ultralight dark matter particle and a standard-model particle, annihilation of two dark matter particles, etc., one needs to carry out the analysis for the range of frequency dispersions between these two limits.

Also, in the case of a decay of a dark matter particle of mass $m_{\chi}$ to a final state of a particle of mass  $m_{\phi}$ and a photon, the photon energy is given by $E_{\gamma} =  \frac{(m_{\chi}^2-m_{\phi}^2) c^2}{2 m_{\chi}}$.  As in the absence of other information concerning $\phi$, the initial mass is unknown, the constraint on the actual decay rate $\lambda_{\phi,\gamma}$ and initial mass $m_{\chi}$ are coupled, $\lambda_{\phi,\gamma} = \lambda_{\gamma,\gamma} \frac{m_{\chi} c^2}{2E_{\gamma}}$, where $\lambda_{\gamma,\gamma}$ is the rate inferred assuming two-photon decay as shown in Eqn. \ref{eqn:Decaypower}.

The future Square Kilometer Array (SKA) will have the benefit of the very large Bose-Einstein enhancement factor associated with the galactic diffuse emission ($2f_{\gamma} \approx 10^{(3-4)}$ for the comparable L-band range as studied here), which could result in sensitivity exceeding current Horizontal Branch Star and CAST limits (\cite{Anastassopoulos2017}).  Whether the analysis techniques explored here, particularly the Doppler asymmetry, and the ability to make collateral use of every spectrum to be collected on the SKA could provide further improvement to their limits will require further study.

In summary, we have carried out a sensitive and selective model-independent analysis technique for dark matter constituting our galactic halo, based on only the most general assumed characteristics of its phase space structure. Signals that are both weak and broad ($\approx mJy, \frac{\Delta\nu}{\nu} \approx 10^{-3}$) can be readily detected at high confidence level. It should be noted that, while the technique here may not find any signature of dark matter, it may prove to be of utility in revealing weak conventional sources associated with the galactic disk.

This analysis will be carried out over the C and X bands in the future.

\section{Acknowledgements}
We gratefully acknowledge the support of the Heising-Simons Foundation, grants 2018-0989 and 2022-3566, and the assistance of Andrew Siemion, Steve Croft, and Matt Lebofsky of the Breakthrough Listen program throughout the project. The Breakthrough Prize Foundation funds the Breakthrough Initiatives which manages Breakthrough Listen. The Green Bank Observatory facility is supported by the National Science Foundation and is operated by Associated Universities, Inc. under a cooperative agreement.  We further thank Ben Safdi, Josh Foster, and Hitoshi Murayama for helpful discussions.  A.K. acknowledges receipt of a graduate research grant from the Applied Science and Technology program.

\newpage
\bibliography{references}{}
\bibliographystyle{aasjournal}

\end{document}